\documentclass{iopjournal}
\usepackage{amsmath}
\usepackage{amssymb}
\usepackage{graphicx}
\usepackage{bm}
\usepackage{booktabs}
\usepackage[section]{placeins}
\usepackage{flafter}
\graphicspath{{}{figs/}}

\begin{document}

\articletype{Paper}

\title{Neutron-source fidelity for laser-driven D--D lithium-blanket
tritium-breeding tests}

\author{Cheng-Qi Zhang$^1$, Yang He$^2$, Mamat Ali Bake$^2$ and Bai-Song Xie$^{1,a)}$}

\affil{$^1$Key Laboratory of Beam Technology of the Ministry of Education, and School of
Physics and Astronomy, Beijing Normal University, Beijing 100875, China}

\affil{$^2$Xinjiang Key Laboratory of Solid State Physics and Devices, School of Physics
Science and Technology, Xinjiang University, Urumqi 830017, China}

\begingroup
\renewcommand{\thefootnote}{a)}
\footnotetext[1]{Author to whom correspondence should be addressed: bsxie@bnu.edu.cn.}
\endgroup

\begin{abstract}
Compact deuterium--deuterium (D--D) neutron sources can provide controllable irradiation fields for
lithium-blanket studies, although their broad joint energy and angle distributions differ from the
conventional $2.45$~MeV isotropic representation. We couple particle-in-cell (PIC) simulations of
target-normal-sheath-accelerated deuterons with a thick-target $D(d,n)^{3}$He source model and Monte
Carlo neutron transport. For natural lithium, the seven two-dimensional sources change tritium
production per source neutron by $-2.5\%$ to $+54.1\%$ relative to the ideal source. The matched
three-dimensional calculation gives an increase of $43.5\%$ and lowers the corresponding
ratio from $1.5406$ to $1.4350$. Source substitutions show that the difference is predominantly
spectral, since the real spectrum alone gives a factor of $1.4199$, while using the real neutron
emission directions in place of isotropic emission adds only a further factor of $1.0106$ in the
three-dimensional case. The real
spectrum lowers the $^{6}$Li
contribution by $6.9\%$, but the accessible $^{7}$Li$(n,Xt)$ response exceeds this loss. Enrichment
to $90\%$ $^{6}$Li keeps the total change within $\pm1.5\%$. In the matched three-dimensional
converter and blanket calculation, direct $D(d,p)$T production is $0.8458$ tritons per source
neutron and accounts for $98.1\%$ and $86.9\%$ of the combined production for natural and enriched
lithium, respectively. High-density polyethylene moderation raises tritium production by about one
order of magnitude but first weakens and then reverses the increase in blanket tritium production. The
analysis quantifies source-model effects in compact breeding tests.
\end{abstract}

\section{Introduction}

Experiments at the National Ignition Facility have exceeded the Lawson criterion for
ignition~\cite{AbuShawareb2022}. Translating such plasma-physics progress into power production also
requires closure of the tritium fuel cycle. A deuterium--tritium plant must supply its own tritium
because natural and external supplies are limited~\cite{Kovari2018,Pearson2018}. Fuel self-sufficiency requires a
lithium breeding blanket with a tritium breeding ratio above unity. Once decay, processing
inventory, reserve, and availability are included, system studies place the achievable margin in a
narrow range of about $1.05$ to $1.15$~\cite{Abdou2021}. Percent-level neutronic biases can
therefore consume a meaningful part of the available margin~\cite{Abdou2015}.

Experimental qualification of breeding blankets remains incomplete, leaving a gap between
conceptual design and an operational fuel cycle~\cite{Federici2023,Gilbert2026}. Accelerator and electrostatic generators can
provide monoenergetic $14.1$~MeV deuterium--tritium or $2.45$~MeV deuterium--deuterium fields, while
ITER test blanket modules are intended to test breeding under reactor-relevant
conditions~\cite{Federici2017,Boccaccini2016,Zinkle2013}. Compact fusion devices provide another
route. Tritium production has been measured in lithium mock-ups driven by compact fusion neutron
sources~\cite{Mukai2021}. A mock-up irradiated during the JET DTE2 campaign has also been compared
with calculation, giving a reported calculation-to-experiment ratio of about
$0.77$~\cite{Fonnesu2024}. Dedicated programmes such as LIBRTI are extending this work toward
integrated irradiation and model validation~\cite{Gilbert2026}. Laser-driven inertial fusion follows
the same deuterium--tritium fuel cycle and will expose its blanket to a pulsed, spectrally broad
neutron field~\cite{Betti2016}.

Short-pulse lasers offer another route to compact neutron generation. Laser-accelerated deuterons
have generated intense, forward-directed neutron fields with broad energy
spectra~\cite{Roth2013,Pomerantz2014,Kar2016,Zulick2013,Alejo2017,Osvay2024,Ditmire1999,Mirfayzi2017,Daido2012}.
Huang \emph{et al.} combined particle-in-cell (PIC) and Monte Carlo calculations for a
collisionless-shock deuteron source and a beryllium breakup converter~\cite{Huang2022,Serber1947}.
Collisionless-shock and relativistic-transparency regimes can produce more energetic or quasi-monoenergetic
ion beams~\cite{Fiuza2012,Haberberger2012,Palaniyappan2015}, while target-normal sheath acceleration
(TNSA) has been characterized to high ion energies~\cite{Fuchs2006,Wagner2016,Higginson2018}. We
apply this cross-scale approach to TNSA deuterons from a solid CD$_2$ foil followed by D--D fusion in
a thick CD$_2$ converter. The resulting neutron field has broad, correlated energy and angle
distributions. Its high-energy tail can access thresholded $^{7}$Li tritium production that is absent
for a $2.45$~MeV D--D source. This source-characterized field allows the effect of source fidelity to
be isolated, although it does not reproduce the $14.1$~MeV neutron field of a D--T reactor or an
inertial-fusion-energy chamber~\cite{Betti2016}.

We quantify the change in inferred blanket TPR when a laser-driven deuterium--deuterium source is
represented by a $2.45$~MeV isotropic source. The calculation couples PIC simulation, a
semi-analytic deuterium--deuterium converter model, and Monte Carlo neutron transport. Source
substitutions isolate the spectral and angular contributions, and isotope-resolved tallies identify
the lithium response. One matched three-dimensional calculation provides a dimensionality
comparison for the $a_0=20$ 2D result. We also compare blanket production with direct $D(d,p)$T production in the
converter and examine how hydrogenous moderation changes both production and source fidelity. All
tritium-production and fidelity quantities are reported per source neutron, while absolute per-shot
yield and tritium recovery remain outside the present scope.

\section{Methods and computational setup}

\begin{figure}[t]
\centering
\includegraphics[width=0.98\linewidth]{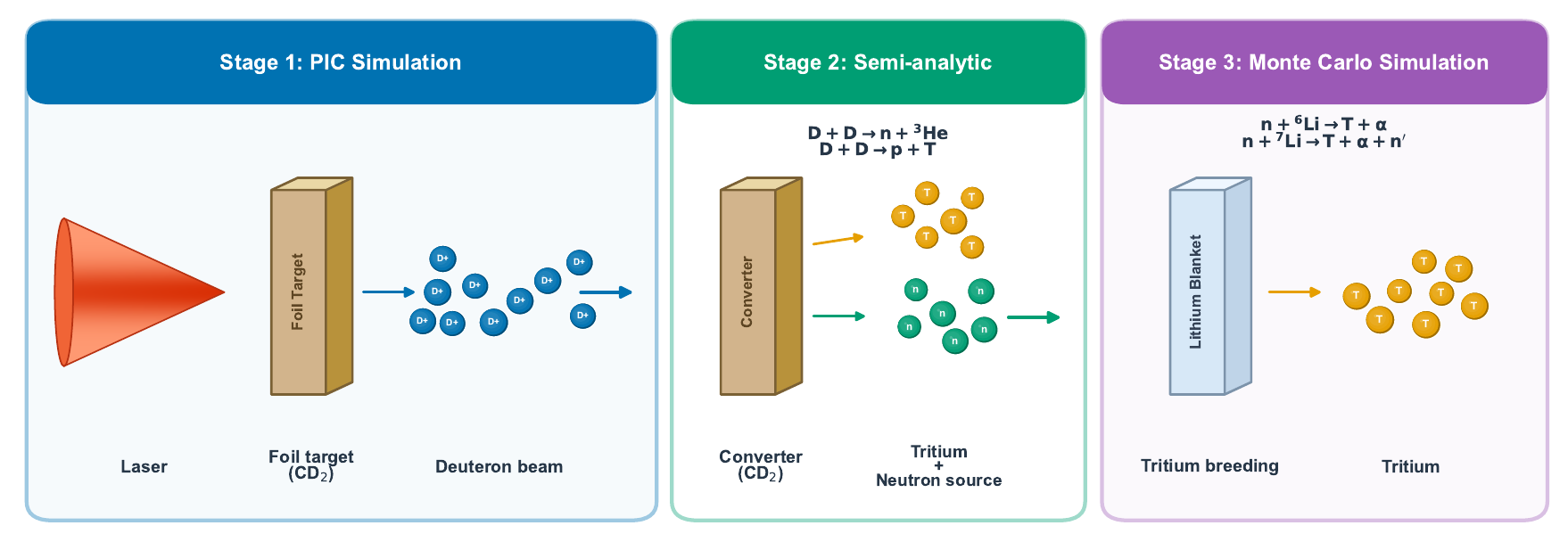}
\caption{Cross-scale calculation. Stage~1 models TNSA deuteron generation from a thin CD$_2$
foil. Stage~2 maps the weighted deuteron beam to thick-target $D(d,n)^{3}$He neutron emission and
the competing $D(d,p)$T branch in a CD$_2$ converter. Stage~3 collapses the converter neutron
emission to a centred point source and uses OpenMC to transport neutrons and score isotope-resolved
tritium production in lithium. }
\label{fig:schematic}
\end{figure}

Figure~\ref{fig:schematic} summarizes the one-way coupled calculation. Stage~1 uses PIC simulation
to produce a weighted deuteron phase-space list, Stage~2 maps the list to $D(d,n)^{3}$He neutrons and
direct $D(d,p)$T production in a converter, and Stage~3 transports the neutrons through lithium
with Monte Carlo methods. For a given neutron-source model, we define the diagnostic tritium
production per source neutron as
\begin{equation}
R \equiv \frac{N_{T,6}+N_{T,7}}{N_{n,\mathrm{src}}},
\label{eq:tpr}
\end{equation}
where $N_{T,6}$ and $N_{T,7}$ are the $^{6}$Li- and $^{7}$Li-born tritons scored in the blanket.  

Each PIC-derived source is transported in three representations. The reference is monoenergetic at
$2.45$~MeV and isotropic, with TPR $R_0$. The spectrum-only representation retains the real energy
spectrum but emits isotropically, with TPR $R_{\mathrm{iso}}$, while the correlated representation
retains the binned joint energy and angle distribution, with TPR $R$. We define the fidelity ratio as
\begin{equation}
F \equiv \frac{R}{R_0}
= \frac{R_{\mathrm{iso}}}{R_0}\,\frac{R}{R_{\mathrm{iso}}}.
\label{eq:closure}
\end{equation}
This exact factorization separates the spectral factor $R_{\mathrm{iso}}/R_0$ from the angular
factor $R/R_{\mathrm{iso}}$. Each source is transported in natural lithium
($7.59$~at.\% $^{6}$Li) and in lithium enriched to $90$~at.\% $^{6}$Li.

\subsection{Particle-in-cell method for the deuteron source}

\begin{figure}[t]
\centering
\includegraphics[width=0.72\linewidth]{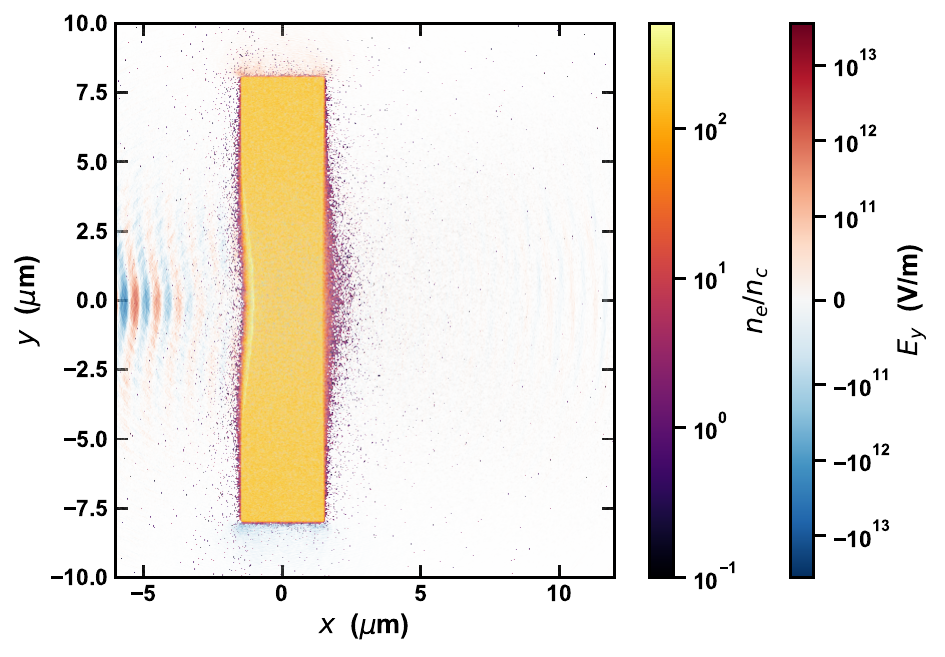}
\caption{Representative Stage~1 interaction at $160$~fs for $a_0=20$ and $t=3~\mu$m. The
electron density $n_e/n_c$ is shown together with the transverse laser field $E_y$ during the
interaction with the overdense CD$_2$ foil.}
\label{fig:laser}
\end{figure}

We model the laser--foil interaction with the relativistic electromagnetic PIC code
EPOCH~\cite{Arber2015}, which advances particles and fields explicitly with a charge-conserving
current deposition~\cite{Birdsall1991,Esirkepov2001}. The linearly polarized Gaussian laser has wavelength
$\lambda_0=0.8~\mu$m, focal radius $w_0=3~\mu$m, duration $\tau=30$~fs, and normalized amplitude
$a_0=5$, 10, 15, or 20. It is normally incident on a solid CD$_2$ foil with electron density
$n_e=3.184\times10^{29}$~m$^{-3}\simeq185n_c$ and a thickness between $1$ and $4~\mu$m. Hot
electrons establish a rear-surface sheath that accelerates deuterons through
TNSA~\cite{Snavely2000,Hatchett2000,Wilks2001,Mora2003,Fuchs2006,Passoni2010,Macchi2013}.
Figure~\ref{fig:laser} shows a representative snapshot of this interaction at $160$~fs for the
$a_0=20$, $t=3~\mu$m case, where the transverse laser field $E_y$ is stopped at the overdense foil
and the electron density $n_e/n_c$ marks the compressed front that drives the sheath.

The 2D scan contains an intensity series with $a_0=5$, 10, 15, and 20 at $t=3~\mu$m and a
thickness series with $t=1$, 2, 3, and $4~\mu$m at $a_0=10$, giving seven unique cases. These
calculations use a $32\times20~\mu$m$^2$ domain ($-6<x<26~\mu$m and $|y|<10~\mu$m), spacings
$\Delta x=16$~nm and $\Delta y=40$~nm, and 16, 32, and 4 macroparticles per cell for electrons,
deuterons, and carbon ions, respectively. Deuterons crossing a plane $10~\mu$m behind the rear
surface are recorded when $E_D>0.4$~MeV and are integrated over time. For assessing convergence,
each recorded deuteron is weighted by the D--D fusion yield it would produce in the Stage~2
converter. Every case is evolved to $6$~ps, at which time the final
$250$~fs interval adds less than $0.2\%$ of this cumulative yield-weighted source, indicating that
its neutron-relevant part has converged. The weighted energy and direction distributions then
provide the input to Stage~2.

\subsection{Semi-analytic method for the D--D converter}

The converter is stationary CD$_2$ with density $1.06$~g\,cm$^{-3}$, corresponding to a deuteron
number density $n_D=7.96\times10^{22}$~cm$^{-3}$
($7.96\times10^{28}$~m$^{-3}$). For a deuteron entering with energy $E_0$, straight-line slowing
to rest gives the reaction yield
\begin{equation}
Y(E_0)=\int_0^{E_0}\frac{n_D\,\sigma(E)}{S(E)}\,\mathrm{d}E ,
\label{eq:thicktarget}
\end{equation}
where $\sigma$ is the D--D cross section and $S$ is the deuteron stopping power. The model assumes that the converter exceeds the continuous-slowing-down range of each incident deuteron, so Eq.~(\ref{eq:thicktarget}) covers the full trajectory from $E_0$ to rest. With the adopted equal-velocity PSTAR proxy, most of the retained deuteron weight has an equivalent range of several to a few tens of micrometres. The range
reaches approximately $0.91$~mm at the maximum 3D energy of $12.4$~MeV and $1.98$~mm at the
maximum retained 2D energy of $19.1$~MeV. These proxy ranges specify the thick-target assumption. We implement the Bosch--Hale parameterizations for
$D(d,n)^{3}$He and $D(d,p)$T~\cite{BoschHale1992}. Their low-energy branching behaviour is checked
against the published data~\cite{BrownJarmie1990}. The parameterizations
are evaluated over their stated centre-of-mass interval
$0.5\leq E_{\mathrm{cm}}\leq4900$~keV, with both numerical cross sections set to zero outside this
interval. Deuteron stopping is represented by NIST PSTAR proton-in-polyethylene data
evaluated at equal velocity and scaled to deuterons~\cite{PSTAR}. 

The reaction-energy probability density is proportional to $\sigma(E)/S(E)$. Each weighted
Stage~1 deuteron record with energy $E_0$, direction $\hat{\bm{\Omega}}_D$, and statistical weight
$w_D$ is mapped to one representative neutron phase-space record. A reaction energy is sampled from
the corresponding cumulative distribution, and isotropic centre-of-mass emission is boosted along
$\hat{\bm{\Omega}}_D$ to obtain the laboratory neutron energy and direction. The neutron record is
assigned statistical weight
\begin{equation}
w_n=w_DY(E_0).
\label{eq:sourceweight}
\end{equation}
The collection of weighted neutron records forms the source passed to Stage~3. This one-to-one
record mapping is a sampling construction. Target deuterons are stationary in the model, while energy straggling, multiple scattering, and transverse converter transport are neglected. The resulting asymmetric high-energy tail is produced by beam-target kinematics, and thermal Doppler broadening is not included. Direct $D(d,p)$T production is evaluated separately with the same stopping model and grid.

\subsection{Monte Carlo method for neutron transport}

OpenMC~\cite{Romano2015} is used in stage~3 to simulate the Neutron transport. The Stage~2 source retains neutron energy,
direction, and statistical weight in the converter. Stage~3
collapses the converter emission to a centred point source inside a spherical cavity of radius
$1$~cm. The surrounding lithium cylinder has radius $10$~cm, height $20$~cm, density
$0.534$~g\,cm$^{-3}$, and vacuum outer boundaries. The correlated model divides
$\mu=\hat{\bm{\Omega}}\mathbin{\cdot}\hat{\bm{z}}$, where $\hat{\bm{\Omega}}$ is the neutron flight direction and $\hat{\bm{z}}$ is the beam axis, into 15 equal bins and constructs a 100-bin conditional energy spectrum in each. Within each bin, $\mu$ and the azimuth are sampled uniformly,
with the azimuth spanning $0$ to $2\pi$. This axisymmetric reconstruction preserves the resolved
joint distribution in $E$ and $\mu$.

The main unmoderated calculations use $100$ batches of $10^6$ source histories, giving $10^8$
histories per run. The moderator calculations use between $10^7$ and $10^8$ histories, depending
on the source and thickness. All production calculations use ENDF/B-VII.1
data~\cite{Chadwick2011}. Tritium is
scored with the OpenMC H3-production tally
and separated by parent isotope. We denote the $^{7}$Li score as $^{7}$Li$(n,Xt)$. Over the energy
range of interest, it is dominated by $^{7}$Li$(n,n't)\alpha$, whose physical laboratory threshold
is $2.819$~MeV. The transport tally uses the tabulated MT205 cross section to ensure the library-consistent. To order the sources by their high-energy content, we use the source-weighted
descriptor
\begin{equation}
\left\langle\sigma_7\right\rangle
=\frac{\sum_i w_i\sigma_{^{7}\mathrm{Li},\mathrm{MT205}}(E_i)}{\sum_i w_i}.
\label{eq:sigma7}
\end{equation}

For the moderation study, a spherical high-density polyethylene (HDPE) shell with density
$0.95$~g\,cm$^{-3}$ and a thickness between $0$ and $5$~cm surrounds the source cavity. The
reference model gives $R_0=0.01118995$ in natural lithium and $0.12832392$ in enriched lithium,
with zero $^{7}$Li score. The implementation is verified against analytic two-body limits, and the
factorization in Eq.~(\ref{eq:closure}) closes to machine precision. The processed MT205 cross
sections from ENDF/B-VIII.0~\cite{Brown2018}, FENDL-3.2~\cite{Schnabel2024}, and
JEFF-3.3~\cite{Plompen2020} agree with the ENDF/B-VII.1 evaluation used here over the relevant
interval.

\section{Results and Discussion}

\begin{figure}[t]
\centering
\includegraphics[width=\linewidth]{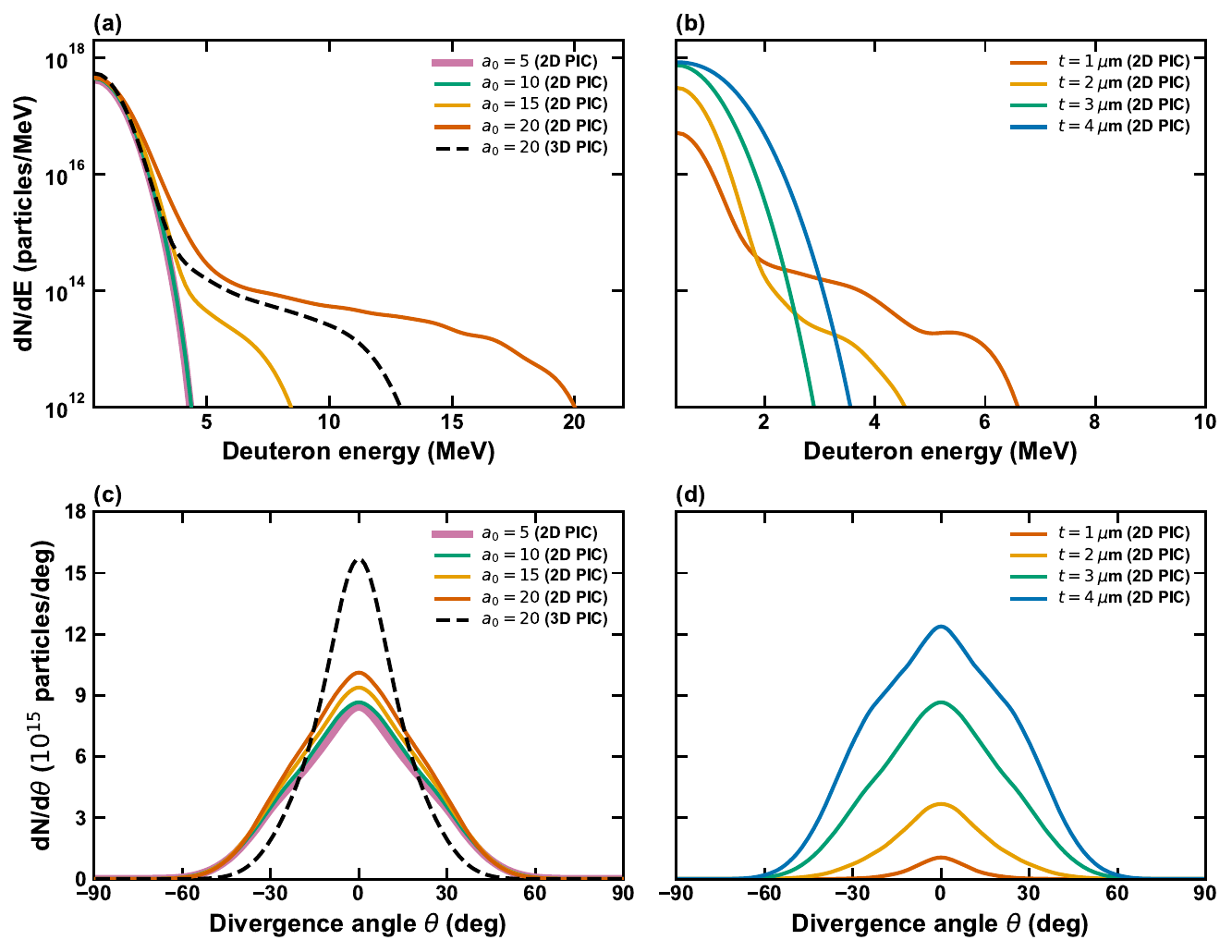}
\caption{Stage~1 deuteron energy spectra in panels (a) and (b) and projected
divergence-angle distributions in panels (c) and (d). The left-hand panels show the
intensity scan at $t=3~\mu$m, while the right-hand panels show the foil-thickness scan at
$a_0=10$. The dashed curves represent the matched 3D calculation at $a_0=20$ and
$t=3~\mu$m.}
\label{fig:deuterons}
\end{figure}

The Stage~1 deuteron distributions are broad and forward directed, as shown in
Fig.~\ref{fig:deuterons}. Along the intensity scan at $t=3~\mu$m, the mean deuteron energy
increases only modestly, from $0.681$~MeV at $a_0=5$ to $0.845$~MeV at $a_0=20$, whereas
the high-energy tail extends to $19.1$~MeV at the highest intensity. Increasing the laser
amplitude therefore affects the energetic tail more strongly than the mean energy. The
corresponding 3D calculation at $a_0=20$ and $t=3~\mu$m produces a lower-energy spectrum, with a
mean energy of $0.549$~MeV and a maximum energy of $12.4$~MeV. The energy response to foil
thickness at $a_0=10$ is non-monotonic. The $1~\mu$m foil produces the most energetic tail, which
extends to $6.0$~MeV, whereas the spectrum from the $3~\mu$m foil ends below $2.2$~MeV.
The mean energy ranges from $0.544$~MeV for the $2~\mu$m foil to $0.882$~MeV for the
$4~\mu$m foil.

Panels~(c) and (d) show the corresponding divergence-angle distributions. In the 2D
calculations, the divergence angle $\theta$ is measured relative to the laser axis within the
simulation plane. The results show that increasing $a_0$ produces only a slight narrowing of the beam, with the RMS
divergence decreasing from $21^\circ$ at $a_0=5$ to $20^\circ$ at $a_0=20$. The dependence
on foil thickness is stronger. At $a_0=10$, the RMS divergence increases from
$10^\circ$ for the $1~\mu$m foil to $23^\circ$ for the $4~\mu$m foil. The angular width is
therefore governed mainly by foil thickness over the parameter range considered, whereas the
high-energy tail is more sensitive to the laser amplitude.

For the matched 3D calculation at $a_0=20$ and $t=3~\mu$m, each deuteron has a full
three-dimensional momentum vector. To obtain a measure comparable with the planar 2D result,
we evaluate the projected divergence angle in each of the two orthogonal planes containing the
laser axis and combine the two distributions. The resulting RMS divergence is $14.5^\circ$,
compared with $20^\circ$ for the matched 2D source. The 3D calculation thus predicts a lower-energy
but more tightly collimated deuteron source under the same nominal laser and target conditions.

\begin{figure}[t]
\centering
\includegraphics[width=\linewidth]{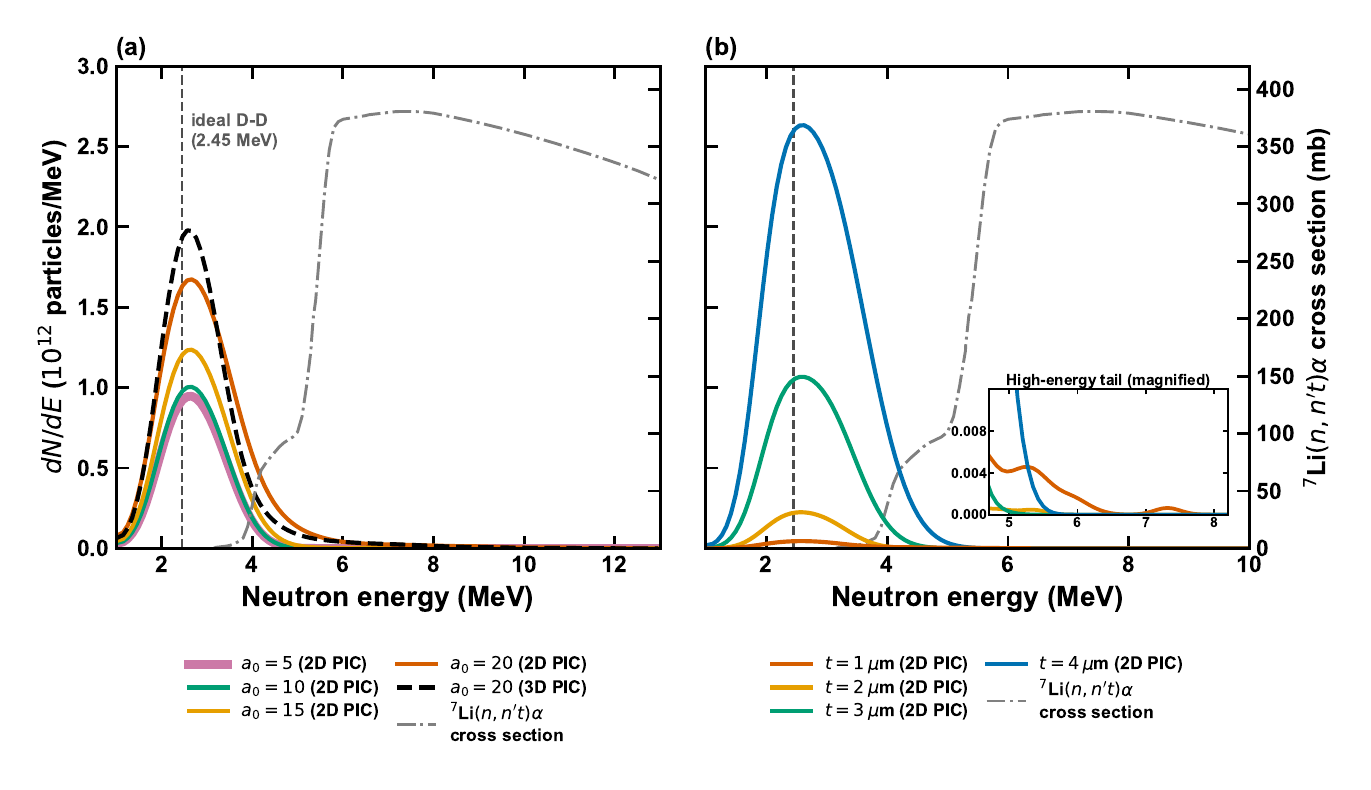}
\caption{Stage~2 neutron energy spectra for (a) the intensity scan and (b) the
foil-thickness scan. The 2D spectra retain their Stage~2 source weights and are not normalized
curve by curve. Because the absolute 2D and 3D PIC weights are not directly comparable, the matched
3D spectrum is rescaled to the integral of the matched 2D spectrum for shape comparison. The
$2.45$~MeV D--D line and the processed $^{7}$Li MT205 tritium-production cross section are
overlaid. The inset in panel~(b) magnifies the low-amplitude high-energy tails on the same
absolute ordinate, using a narrower smoothing kernel to resolve their extent.}
\label{fig:neutron}
\end{figure}

Figure~\ref{fig:neutron} shows how the converter maps the deuteron distributions into
asymmetric neutron spectra around the D--D peak. Every source peaks near the $2.45$~MeV D--D line,
while the spectra differ in the forward-boosted tail above it. Along the intensity scan the mean neutron energy increases from
$2.73$~MeV at $a_0=5$ to $3.04$~MeV at $a_0=20$. In the thickness scan it is also non-monotonic, with
the $2~\mu$m source having the lowest mean neutron energy at $2.66$~MeV and the $1~\mu$m source the highest at $2.97$~MeV,
following the high-energy deuteron content. The magnified tail in Fig.~\ref{fig:neutron}(b)
further shows that the $1~\mu$m spectrum extends to higher neutron energies than the other
thickness-scan spectra. The matched 2D and 3D sources have mean neutron energies of $3.044$ and $2.906$~MeV and source-weighted $\langle\sigma_7\rangle$ of $19.3\times10^{-3}$ and
$15.4\times10^{-3}$~b, respectively. The reference $2.45$~MeV source lies below the $^{7}$Li$(n,Xt)$ threshold and therefore
produces no $^{7}$Li tritium in the evaluated data, while the real sources extend above this threshold. In beam-target D--D reactions, the centre-of-mass frame moves along the incident-deuteron direction. Neutrons emitted in this direction are shifted to higher energies in the laboratory frame, while those emitted in the opposite direction are shifted to lower energies. This kinematic broadening produces a high-energy wing above the nominal $2.45$~MeV line and thereby opens the $^{7}$Li$(n,Xt)$ channel that is inaccessible to the reference source.

\begin{figure}[t]
\centering
\includegraphics[width=\linewidth]{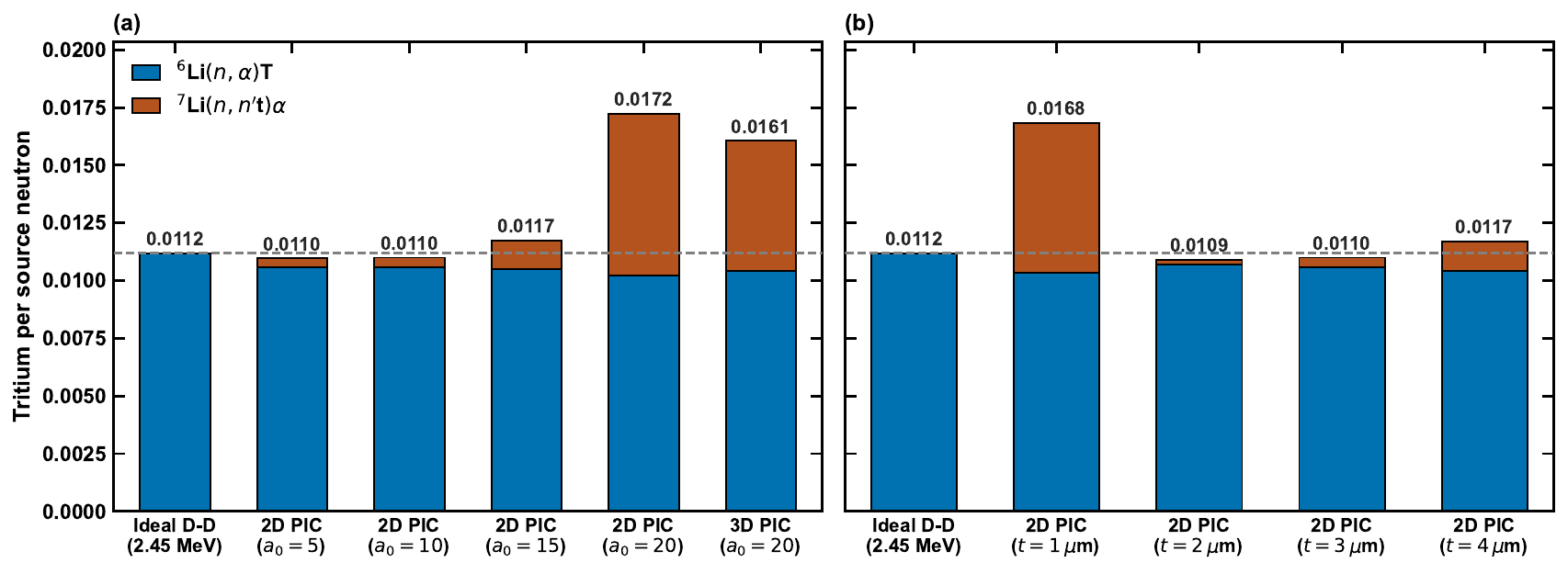}
\caption{Tritium production per source neutron in natural lithium, split into
$^{6}$Li$(n,\alpha)$T (blue) and $^{7}$Li$(n,Xt)$ (orange), for (a) the intensity scan and (b) the
thickness scan. The dashed horizontal line marks the TPR from the $2.45$~MeV isotropic reference
source.}
\label{fig:main}
\end{figure}

Figure~\ref{fig:main} resolves the blanket response into its isotope-specific contributions,
while Table~\ref{tab:results} summarizes the corresponding fidelity ratios. The
$^{6}$Li$(n,\alpha)$T contribution (blue) remains below the reference value of $0.0112$ per
source neutron for every real source and varies relatively little across the source set,
decreasing slightly as spectral weight moves away from the low-energy capture region. By contrast, the $^{7}$Li$(n,Xt)$ contribution is dependent on the neutron
spectrum. It increases from $2.1\times10^{-4}$ per source neutron for the
$a_0=10$, $t=2~\mu$m source to $7.0\times10^{-3}$ for the
$a_0=20$, $t=3~\mu$m source, reflecting the greater spectral overlap with the
$^{7}$Li tritium-production cross section. Consequently, the net tritium production of the actual source relative to the reference is determined by the balance between increased $^7\text{Li}$ production and decreased $^6\text{Li}$ capture.   

For the 2D sources in natural lithium, $F$ ranges from $0.9752$ to $1.5406$. The minimum occurs
for the $a_0=10$, $t=2~\mu$m source because its additional $^{7}$Li contribution does not fully
compensate for the loss of $^{6}$Li production. By contrast, the $a_0=20$, $t=3~\mu$m source
places more spectral weight in the energy range relevant to $^{7}$Li tritium production and
increases the TPR by $54.1\%$ relative to the reference source. The response to foil thickness is
not monotonic either. The $1~\mu$m case has the smallest integrated Stage~2 source weight in
Fig.~\ref{fig:neutron}(b), but its high-energy tail extends beyond those of the other
thickness-scan cases. Table~\ref{tab:results} gives the corresponding source-weighted
cross section as $\langle\sigma_7\rangle=19.1\times10^{-3}$~b, the largest value in the thickness
scan. Because $F$ is defined per source neutron, this spectral overlap rather than the integrated
source weight determines the ordering, and the $1~\mu$m source gives $F=1.5043$. The matched 3D source gives
$R=0.0160570$, compared with $R_0=0.0111900$, and hence
$F=1.43495\pm0.00029$, where the uncertainty is the
one-standard-deviation statistical uncertainty from the Monte Carlo transport calculation. This corresponds to a $43.5\%$ increase over the reference result, smaller than the $54.1\%$ increase obtained from the matched 2D source and consistent with
the lower-energy neutron spectrum of the 3D calculation.

\begin{table}[t]
\centering
\caption{Per-source-neutron blanket fidelity. The descriptor $\langle\sigma_7\rangle$ is the
source-weighted $^{7}$Li$(n,Xt)$ MT205 cross section defined in Eq.~(\ref{eq:sigma7}), used here as
a measure of the high-energy content of the source spectrum. The quantity $F=R/R_0$ is the total TPR ratio of the correlated
real source to the reference source.}
\label{tab:results}
\begin{tabular}{lccc}
\toprule
Source ($a_0$, $t$) & $\langle\sigma_7\rangle$ ($10^{-3}$~b) & $F$ (natural) & $F$ (90\% $^{6}$Li) \\
\midrule
$a_0{=}5$, $3~\mu$m  (2D)          & 1.26 & 0.9797 & 1.0067 \\
$a_0{=}10$, $3~\mu$m (2D)         & 1.41 & 0.9827 & 1.0062 \\
$a_0{=}15$, $3~\mu$m (2D)         & 3.78 & 1.0476 & 1.0021 \\
$a_0{=}20$, $3~\mu$m (2D)     & 19.3 & 1.5406 & 0.9861 \\
$a_0{=}20$, $3~\mu$m (3D)     & 15.4 & 1.4349 & 0.9954 \\
$a_0{=}10$, $1~\mu$m (2D)         & 19.1 & 1.5043 & 0.9910 \\
$a_0{=}10$, $2~\mu$m (2D)        & 0.65 & 0.9752 & 1.0114 \\
$a_0{=}10$, $4~\mu$m (2D)        & 4.15 & 1.0456 & 0.9987 \\
\bottomrule
\end{tabular}
\end{table}

For the 3D anchor, the spectral factor is $R_{\mathrm{iso}}/R_0=1.41985$ and the angular factor is
$R/R_{\mathrm{iso}}=1.01063$. The angular factor is the change from using the real neutron emission
directions instead of isotropic emission, and it moves the result by only $1.06\%$. Across all sources considered, the
angular factor changes TPR by no more than $1.35\%$. This upper bound applies only to the centred,
approximately symmetric point-source geometry and should not be extrapolated to an external or
strongly asymmetric irradiation configuration. In the isotope-resolved 3D tally, the real source
lowers the $^{6}$Li term from $0.0111900$ to $0.0104188$ per source neutron, a
$6.9\%$ reduction, while adding $0.0056382$ through $^{7}$Li MT205. The $^{7}$Li contribution
exceeds the $^{6}$Li loss in the positive cases, whereas the sources with smaller high-energy tails retain a
net loss. With $90\%$ $^{6}$Li, the ratios remain within $\pm1.5\%$ of unity and the 3D value is
$0.99543$.

The source-weighted $\langle\sigma_7\rangle$ listed in Table~\ref{tab:results} orders the
natural-lithium fidelity ratios, which increase nearly linearly with it across the source set, so it
serves as a compact hardness descriptor rather than a precise predictor. The dimensionality of the
source nevertheless matters at the matched point. The ratio $F$ decreases by $6.9\%$, from $1.5406$
in 2D to $1.4350$ in 3D. Expressed as the enhancement above the reference, the 2D source gives $54.1\%$
and the 3D source $43.5\%$, so the 2D calculation overstates the enhancement by $24\%$.

\begin{figure}[t]
\centering
\includegraphics[width=\linewidth]{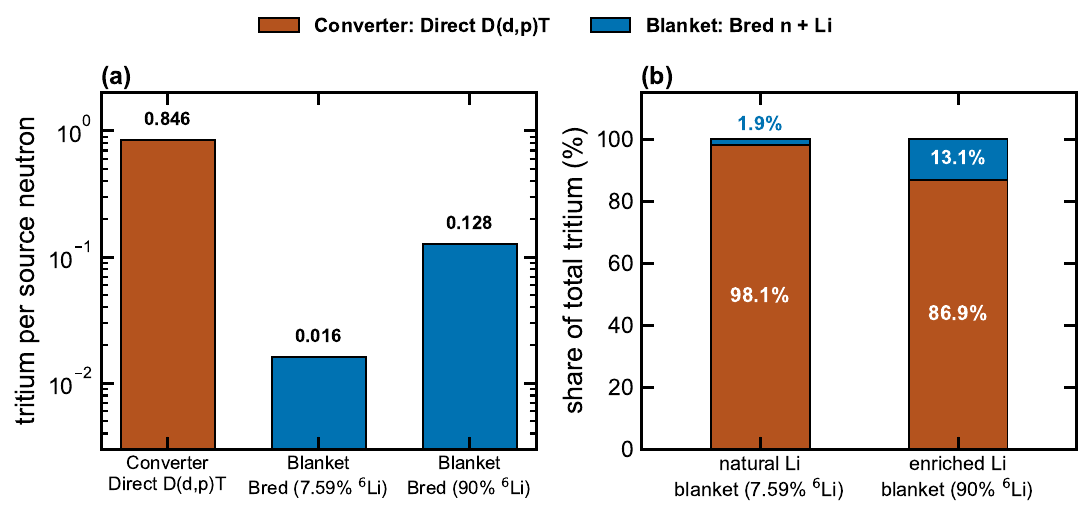}
\caption{Normalized tritium-production balance for the 3D anchor. Panel (a) compares
converter-born $D(d,p)$T with blanket-born tritium per source neutron. Panel (b) shows the
fractional contributions for natural and $90\%$-$^{6}$Li blankets. Tritium transport, retention,
release, and recovery are not included.}
\label{fig:budget}
\end{figure}

Blanket-born tritium is only one part of the total tritium produced by the system
(Fig.~\ref{fig:budget}). The CD$_2$ converter also produces tritium directly, through the $D(d,p)$T
branch of the same D--D fusion reactions that generate the neutrons, with no blanket involved. This
direct channel gives $0.845805$ tritons per source neutron, compared with $0.016057$ bred in the
natural-lithium blanket and $0.127738$ in the enriched blanket, so it accounts for $98.1\%$ and
$86.9\%$ of the combined tritium and exceeds the natural-lithium blanket term by a factor of $52.7$.
Because the converter makes about fifty times more tritium than the natural-lithium blanket breeds,
an experiment that recovers tritium without physically separating the converter from the blanket
would be dominated by this direct $D(d,p)$T contribution. This dominance also dilutes the source-fidelity signal. The direct converter term is identical for
the ideal and the real neutron source, so when it is added to both budgets the total
natural-lithium tritium changes by only $0.57\%$ on replacing the ideal source with the real one. The source-fidelity effect is therefore
observable only if the blanket-bred tritium is measured separately from the converter-born tritium.
Because the converter and the blanket are physically distinct components whose tritium is recovered
separately, the blanket-specific change of $43.5\%$, and not the diluted combined value, is the
quantity relevant to a breeding measurement.

\begin{figure}[t]
\centering
\includegraphics[width=\linewidth]{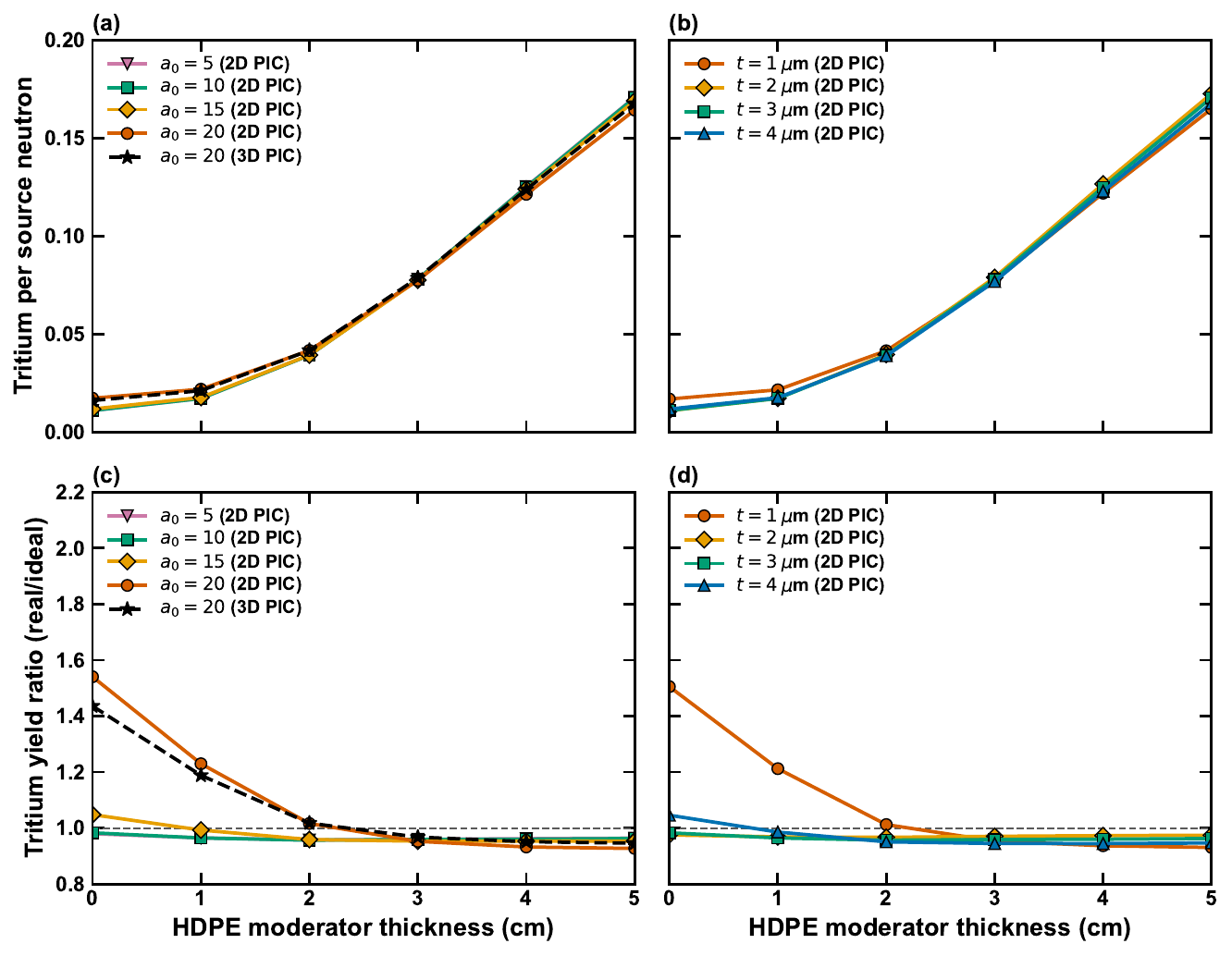}
\caption{Natural-lithium response to a spherical HDPE shell of density $0.95$~g\,cm$^{-3}$.
(a,b) Tritium production per source neutron and (c,d) $F$ versus moderator thickness for the
intensity and target-thickness scans. The black dashed, star-marked curve in panels (a) and (c) is
the matched 3D anchor.}
\label{fig:moderator}
\end{figure}

The $^{6}$Li$(n,\alpha)$T cross section rises steeply toward low neutron energy, so slowing the fast
D--D neutrons in a hydrogenous medium moves them into this high-capture region and raises tritium
production. Capture of such lower-energy neutrons by $^{6}$Li is the dominant breeding reaction in
most blanket concepts. We therefore surround the source with a high-density polyethylene (HDPE)
shell and examine how moderation changes both the tritium yield and the source-fidelity effect as exhibited in fig.~\ref{fig:moderator}. We sweep the HDPE thickness over $0$ to $5$~cm. For the
$a_0=20$ 3D anchor, the natural-lithium TPR rises from $0.01607$ per source neutron with no shell to
$0.04169$ at $2$~cm and $0.16772$ at $5$~cm. The value at $5$~cm
is larger by a factor of $10.4$ because thermalization shifts the response toward the large
low-energy $^{6}$Li capture cross section. Over the same thicknesses, $F$ falls from $1.4357$ to
$1.0174$ and then $0.9457$. Moderation therefore weakens and reverses the enhancement caused by the source's high-energy tail.

The present results should be interpreted as a source-fidelity assessment rather than an
absolute prediction of tritium yield per laser shot. The numerical TPR values are specific to
the compact lithium cylinder and centred point-source geometry considered here. The converter
is represented in the thick-target limit using equal-velocity PSTAR stopping, and the neutron
field is reconstructed from an axisymmetric binned $E$--$\mu$ distribution. These approximations
affect the quantitative response but do not change the spectral-threshold mechanism identified
above. In particular, restricting the Bosch--Hale cross sections to their stated energy range
and using the processed MT205 data near threshold both suppress the calculated high-energy
contribution, so the reported $^{7}$Li response is conservative within the adopted model. The
converter-born $D(d,p)$T term should likewise be regarded as a normalized production term because
tritium transport, retention, release, and recovery are not included. Future work will replace
the stopping-power proxy with material-specific deuteron transport, retain the finite spatial
and angular extent of the converter source, and extend the matched comparison to a broader set
of 3D PIC conditions. Coupling these developments to measured neutron spectra and blanket
tritium yields will establish how the per-source-neutron fidelity correction obtained here
transfers to a specified experimental converter--blanket assembly.
\section{Conclusion}

We coupled PIC simulation, a thick-target converter model, and neutron transport to examine
source-model effects in a compact D--D lithium-blanket test. For natural lithium, the seven 2D sources give $F$ values
from $0.975$ to $1.541$, while the matched $a_0=20$ 3D calculation gives $F=1.435$. The
factorization shows that the real spectrum drives most of this change by reducing $^{6}$Li
production and opening a larger $^{7}$Li$(n,Xt)$ contribution. Using the real emission directions in place of isotropic emission changes the centred-geometry
result by about $1\%$, and enrichment to $90\%$ $^{6}$Li keeps the total source effect within
$\pm1.5\%$. Tritium produced through the converter
$D(d,p)$T branch accounts for $98.1\%$ and $86.9\%$ of the normalized combined production for
natural and enriched lithium. Moderation
raises tritium production by about an order of magnitude but first weakens and then reverses the
increase from the high-energy tail, so moderator design must be included when interpreting source
fidelity. A CD$_2$-specific deuteron stopping calculation, a broader set of 3D PIC sources, and an integral tritium-production benchmark will be the main steps to extend the present
per-source-neutron analysis in the future.
\ack{This work is supported by the National Natural Science Foundation of China (NSFC) under Grants No. 12375240 and No. 12535015, and by the Program of China Scholarship Council (Grant CSC202506040219). M. A. Bake was supported by the Special Training Program of the Science and Technology Department of Xinjiang, China (Grant No. 2024D03007).}

\section*{Conflict of Interest}
The authors declare no conflicts of interest.

\data{The data that support the findings of this study are available from the corresponding author upon reasonable request.}

\roles{\textbf{Cheng-Qi Zhang:} Conceptualization (lead); Methodology (lead); Software (lead); Investigation (lead); Formal analysis (lead); Data curation (lead); Writing original draft (lead). \\
\textbf{Yang He:} Methodology (supporting); Writing review \& editing (supporting). \\
\textbf{Mamat Ali Bake:} Formal analysis (supporting); Writing review \& editing (supporting). \\
\textbf{Bai-Song Xie:} Supervision (lead); Funding acquisition (lead); Writing review \& editing (lead).}

\bibliographystyle{unsrt}
\bibliography{ref}

\end{document}